# Crystallography, relativity and octonions

R J Potton

Magdalene College, Cambridge

**Abstract**

In condensed matter theory many invaluable models rely on the possibility of subsuming fundamental particle interactions in constitutive relations for macroscopic fields in near equilibrium assemblies of particles. Should one wish to maintain relativistic covariance this substitution generates a problem that can only be addressed by expanding the dimension of the space-time base manifold (four) to that of its tangent bundle (eight). The linear vector space of the octonions over the real (or conceivably rational) field seems to offer definite advantages in doing this.

rjp73@cam.ac.uk

# 1 Introduction

Instructive examples of the exploitation of constitutive relations among macroscopic fields alluded to in the abstract are to be found in a book by Butcher and Cotter [1]. Discussions of the dimension of the world [2] do not preclude the possibility that what is generally focussed upon may be but one aspect of the mathematical framework that it is appropriate to employ. For example, the observation that the theories of condensed matter physics are neither relativistically covariant nor conformal, adduced as an argument against the usefulness of holographic duality in that arena [3], ignores the possibilities afforded by extending the base manifold of space-time to its tangent bundle [4].

There are a number of other pointers toward a need to regularize the algebraic basis of Maxwell's equations as applied to polarizable and magnetizable media. These include:

(i) Gogberashvili's [5] formulation of Maxwell's equations for free space in terms of octonions;

(ii) Frankel's [6] acknowledgement that electromagnetic constitutive relations are in some sense "complicated";

(iii) the inability of tensor calculus using symmetry on its own to predict the number of independent permittivity and permeability constants for crystals of low symmetry [7,8,9];

(iv) indications that, even at the most primitive level, connections between events may deny algebraic associativity [10] but that Jacobi identity nonassociativity that is well used in the transition from Lie groups to Lie algebras is only one possibility for accommodating this. Further, that the primacy of symmetry groups over their algebraic representations may need to be reconsidered; and

(v) the desirability of clarifying the status of discrete symmetries that are, at one and the same time, involutions and not involutions [11].

The present proposition is that, at their most basic level, electromagnetic constitutive relations arise from the duality of subspaces of octonion space enabled by the existence of the triality [12] that forms part of the alternative nonassociative algebra of octonions [13].

Octonion co-ordinatization of space time makes room for the explicit handling of scalars vectors, bivectors and trivectors [14]. The many possible sign changes of the units of octonion algebra are involutions that engender projections of octonion space into its subspaces in the same way that the involution of space inversion can be used to define projection operators of three-dimensional space that are respectively of even or odd parity. By the same token, there are projections of octonion space that have a special significance in that they achieve a separation, not of space and time, but of space from a two-dimensional representation of space-time. These projections are defined below. The special significance of these projections among all possible others is connected with additional properties of the subspaces that arise. Specifically, the two dimensional subspace is capable of being equipped with a symmetric bilinear form whilst the six-dimension space separated from it is symplectic. Some further discussion of the number theoretic context of these statements is provided below. The six-dimensional space in turn admits of an involution based on the torsion group $Z_2$ being the quotient group $Z_6/Z_3$. This involution gives rise to a separation of the subspace of dimension six into subspaces of three-vector and three-bivector character, the connection being the duality that belongs naturally to projective three-space [15].

In the context of electromagnetism the fundamental field E and flux density B [6] belong respectively to three-vector and three-bivector subspaces whilst the auxiliary variables D and H belong to the same subspaces but in contrary manner. The derivation of flux densities from fields, captured by constitutive relations, depends on the existence of the trivector that is part of the alternative version of nonassociative algebra to which the octonions belong. Upon separation of space from space-time that trivector appears as part of a two-dimensional representation of four-dimensional space-time.

In the absence of crystallographic symmetry the constitutive relation between D and E, as well as that between B and H, contains the octonion trivector. There is room besides for only three independent permeability or permittivity components since, for example, electric polarization orthogonal to the yz-plane (4-plane) is insensitive to fields in the y- and z- (2 and 3) directions (It should be clear that the co-ordinates referred to here are in general affine rather than rectangular Cartesian). This resolves an issue in the tensor theory of dielectric response where, in triclinic crystals, the crystal class on its own predicts a property tensor having six independent components [7]. Using tensor calculus, considerations beyond pure symmetry are required to arrive at the correct number of constants (three) this being the number of real eigenvalues of a symmetric (or possibly Hermitian) matrix [8,9].

## 2 Involutions and their associated projections

The octonions (Cayley numbers) are hypercomplex numbers the units of which obey pairwise multiplication as specified in table 1.

|       | $e_0$ | $e_1$ | $e_2$ | $e_3$ | $e_4$ | $e_5$ | $e_6$ | $e_7$ |
|-------|-------|-------|-------|-------|-------|-------|-------|-------|
| $e_0$ | 1     | $e_1$ | $e_2$ | $e_3$ | $e_4$ | $e_5$ | $e_6$ | $e_7$ |
| $e_1$ | $e_1$ | -1    | $-e_6$| $e_5$ | $e_7$ | $-e_3$| $e_2$ | $-e_4$|
| $e_2$ | $e_2$ | $e_6$ | -1    | $-e_4$| $e_3$ | $e_7$ | $-e_1$| $-e_5$|
| $e_3$ | $e_3$ | $-e_5$| $e_4$ | -1    | $-e_2$| $e_1$ | $e_7$ | $-e_6$|
| $e_4$ | $e_4$ | $-e_7$| $-e_3$| $e_2$ | -1    | $e_6$ | $-e_5$| $e_1$ |
| $e_5$ | $e_5$ | $e_3$ | $-e_7$| $-e_1$| $-e_6$| -1    | $e_4$ | $e_2$ |
| $e_6$ | $e_6$ | $-e_2$| $e_1$ | $-e_7$| $e_5$ | $-e_4$| -1    | $e_3$ |
| $e_7$ | $e_7$ | $e_4$ | $e_5$ | $e_6$ | $-e_1$| $-e_2$| $-e_3$| -1    |

**Table 1** Multiplication table of octonion units

The entries in table 1 are algebraically freely generated by the units $e_1$, $e_2$ and $e_3$. The choice of the field of scalars $\{o_n\}$ for the ensuing linear vector space is discussed in paragraph 3 of section 3 below.

An octonion can be written in terms of the units given in table 1 thus:

$$o = o_0 e_0 + o_1 e_1 + o_2 e_2 + o_3 e_3 + o_4 e_4 + o_5 e_5 + o_6 e_6 + o_7 e_7$$

and may be considered to be composed of parts that have even or odd superparity according as the monomial in $e_1$, $e_2$ and $e_3$ of the respective unit is of even or odd degree.

The associated superinversion operation may be called $\pi$ and is defined by the equation:

$$\pi o = o_0 e_0 - o_1 e_1 - o_2 e_2 - o_3 e_3 + o_4 e_4 + o_5 e_5 + o_6 e_6 - o_7 e_7 \,.$$

It is an involution and gives rise to projection operators for the even and odd parts of an octonion, thus:

$$P_{\pi_+} o = \frac{1+\pi}{2} o = o_0 e_0 + o_4 e_4 + o_5 e_5 + o_6 e_6$$

$$P_{\pi_-} o = \frac{1-\pi}{2} o = o_1 e_1 + o_2 e_2 + o_3 e_3 + o_7 e_7$$

Another involution operation which we may call **ρ** (for relativistic) is defined by the equation:

$$\rho o = e_7 o e_7 = -o_0 e_0 + o_1 e_1 + o_2 e_2 + o_3 e_3 + o_4 e_4 + o_5 e_5 + o_6 e_6 - o_7 e_7$$

In this case the projection operators for subspaces of octonion space that are respectively even and odd under **ρ** are $P_{\rho+}$ and $P_{\rho-}$ where:

$$P_{\rho+} o = \frac{1 + e_7 o e_7}{2} = o_1 e_1 + o_2 e_2 + o_3 e_3 + o_4 e_4 + o_5 e_5 + o_6 e_6$$

$$P_{\rho-} o = \frac{1 - e_7 o e_7}{2} = o_0 e_0 + o_7 e_7$$

The separation of the, even under **ρ,** six-dimensional subspace from the two-dimensional relativistic subspace of the relativistic bundle is special because of the different two-forms that they support (symplectic and symmetric bilinear respectively). It is should not be surprising that there is a two-dimensional representation of relativistic space-time [16] since, under a Lorentz boost, volume is contracted by the same factor that time is dilated.

The product of the involutions **π** and **ρ** is also an involution:

$$\pi \rho o = -o_0 e_0 - o_1 e_1 - o_2 e_2 - o_3 e_3 + o_4 e_4 + o_5 e_5 + o_6 e_6 + o_7 e_7$$

**π ρ** is of more than passing interest as, unlike **π** and **ρ**, it is a discrete symmetry operation of the proper Lorentz group of the base manifold.

Based on this inventory of involutions and their associated projection operators it is possible to posit a splitting of octonion space according to the scheme shown in figure 1. However, this proposition begs the question of what makes this particular splitting special among all possible others. To answer this one must look at the algebraic properties of the respective subspaces that are engendered not by the norm that gives the octonions their status as a division algebra but rather by the exterior product that is their most distinctive feature as an exterior (Ausdehnung) algebra. This is the subject of the next section.

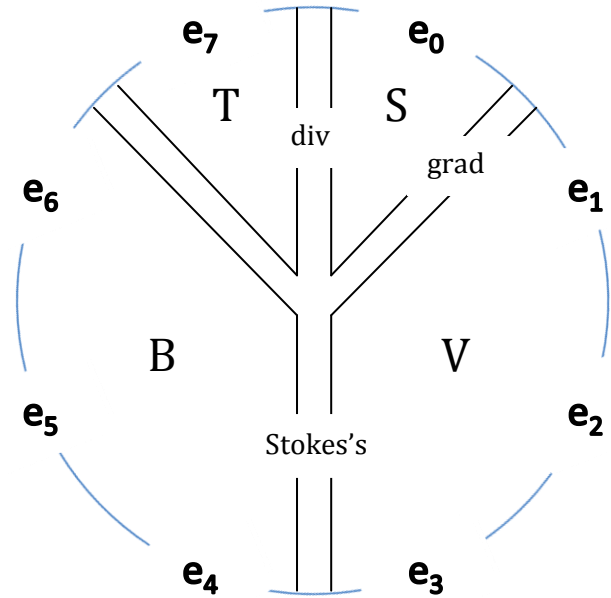

**Figure 1** Showing the leaf spaces connected by integral theorems (divergence, gradient and Stokes's) of the tangent bundle; the octonion units are those that conform to the pairwise multiplication given in table 1; S, T, V and T denote scalar, vector, bivector and trivector elements of the exterior algebra.

Separation of the vector and bivector subspaces is achieved by the products of projection operators $P_{\rho+} P_{\pi-}$ and $P_{\rho+} P_{\pi+}$ respectively.

## 3 Varieties and semi-algebraic sets on projected subspaces

The quadratic form on the subspace $\pi\rho\text{-}$ is:

$$(P_{\pi\rho\_} o)^2 = (o_0 e_0 + o_1 e_1 + o_2 e_2 + o_3 e_3)^2$$
$$= (o_0^2 - o_1^2 - o_2^2 - o_3^2) + 2 o_0(o_1 e_1 + o_2 e_2 + o_3 e_3)$$

Were this form to have only its real part it would be natural to interpret its vanishing as defining an algebraic set (variety) identical to the light cone in the base manifold. It may be considered equally natural to interpret its imaginary part as determining an inequality that, together with another from the real part, defines the interior of the future light cone as a semi-algebraic set.

## 4 Projective geometry

Apart from its role with $\pi$ in the separation of the base manifold from its tangent bundle the involution $\rho$ is deserving of consideration in respect of the $\rho+$ subspace it engenders. $e_7$ terms are absent from this subspace and the relevant part of the Fano plane [12] can take either of the forms shown in figures 2(a) and 2(b) according as $e_4$, $e_5$ and $e_6$ or $e_1$, $e_2$ and $e_3$ are taken as generators of the algebra. The interchanging of these combinations of units constitutes a duality of the six dimensional subspace and is part of the structure of projective three-space [17].

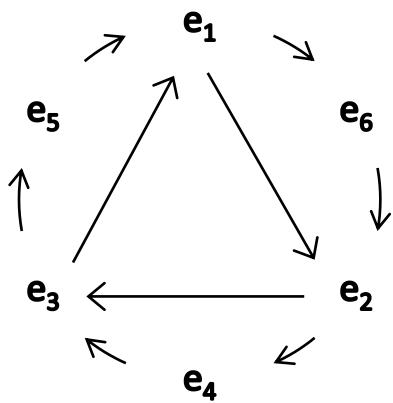 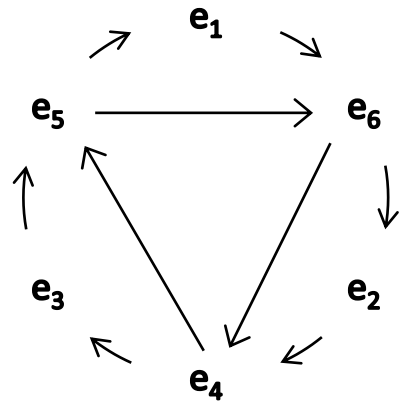

**Figure 2(a)** generators $e_4$, $e_5$ and $e_6$      **Figure 2(b)** generators $e_1$, $e_2$ and $e_3$ as table 1

The separated subspaces identified in section 2 are linear vector spaces over number fields. Working from integer measures for bivectors and trivectors that arise naturally for crystal lattices one sees that the relationships of crystallographic reciprocal space are captured by the octonion product as a rational * [18] symplectic form on the $\rho_+$ subspace. It remains to be determined whether the rational numbers or the real numbers constitute the appropriate field for octonion space as a whole. The answer to this question will depend on consideration of how the analytical properties of the relativistic subspace $\rho$- may be reconciled with those of the dualised projective subspace $\rho$+.

* "Gauss was interested in the question of the rationality or irrationality of the ratios of crystallographic coefficients." Dunnington [19]

## 5 Octonion form of electromagnetic constitutive equations

The general constitutive relations for macroscopic electric and magnetic fields in polarizable media are set out in reference [20]. Written as equations in octonion algebra they are:

$$D_4 \mathbf{e_4} = \varepsilon_0 \mathbf{e_7} E_1 \mathbf{e_1} + P_4 \mathbf{e_4}$$

$$H_1 \mathbf{e_1} = \frac{1}{\mu_0 \mathbf{e_7}} B_4 \mathbf{e_4} - M_1 \mathbf{e_1} = -\frac{\mathbf{e_7}}{\mu_0} M_4 \mathbf{e_4} - M_1 \mathbf{e_1}$$

with four further equations obtained by simultaneous cyclic permutations (1,2,3) and (4,5,6).

In these equations the free space permittivity and permeability, $\varepsilon_0 \mathbf{e_7}$ and $\mu_0 \mathbf{e_7}$, must be placed in the correct position on account of the noncommutative nature of the algebra. These constants acquire some of the attributes of operators and in accordance with this have been placed to the left of the fundamental field * in the expression for the associated derived field. This convention agrees with the form of the octonion multiplication table given above and is connected with the right hand rule that determines the orientation of bivectors from that of the volume form trivector. The latter observation suggests the introduction of a further involution, $\chi$, to represent the transition from right-handed to left-handed association of bivector and trivector orientations:

$$\chi o = o_0 \mathbf{e_0} + o_1 \mathbf{e_1} + o_2 \mathbf{e_2} + o_3 \mathbf{e_3} - o_4 \mathbf{e_4} - o_5 \mathbf{e_5} - o_6 \mathbf{e_6} + o_7 \mathbf{e_7}$$

The implications of the existence of such an involution are explored in section 6 below.

A linear approximation of the polarization and magnetization in terms of the fundamental fields E and B allows electric and magnetic susceptibility constants and relative permeability and permittivity constants to be defined.

$$P_4 \mathbf{e_4} = \chi_4 \varepsilon_0 \mathbf{e_7} E_1 \mathbf{e_1}$$

$$M_1 \mathbf{e_1} = \chi_1 \frac{1}{\mu_0 \mathbf{e_7}} B_4 \mathbf{e_4} = -\chi_1 \frac{\mathbf{e_7}}{\mu_0} B_4 \mathbf{e_4}$$

Once again equations for the other components of polarization and magnetization may be obtained by simultaneously cyclically permuting (1,2,3) and (4,5,6).

Written in octonion form it becomes clear that the macroscopic field vectors belong to a vector space equipped with its own algebraic structure (a symplectic bilinear form).

---

* The fields E and B are fundamental in the sense that they are the spatial averages over a suitable volume of the free space fields between particles.

# 6  Multivector orientations and a $\chi\pi\rho$ theorem

Classic texts, with the exception of Frankel's, have remarkably little to say about the orientation of geometrical objects that are considered to be representative of physical entities but from what has been said here it will be apparent that the product of involutions $\chi$, $\pi$ and $\rho$ has interesting properties in respect of the trivector $o_7\mathbf{e_7}$ :

$$\chi\pi\rho\, o = -o_0\mathbf{e_0} - o_1\mathbf{e_1} - o_2\mathbf{e_2} - o_3\mathbf{e_3} - o_4\mathbf{e_4} - o_5\mathbf{e_5} - o_6\mathbf{e_6} + o_7\mathbf{e_7}.$$

Uniquely among the parts of an octonion $o_7\mathbf{e_7}$ is invariant under $\chi\pi\rho$. In this sense, three-space is oriented whereas the orientations of all other components of the octonion tangent bundle are susceptible to change by the operator $\chi\pi\rho$  In respect of bivectors, contact is made with simplicial homology if $\chi$ is identified as a boundary operator on the three-space simplex measured by the trivector $o_7\mathbf{e_7}$ ; for scalars and vectors, products of $\chi$ with $\rho$ and $\pi$ respectively are required to produce changes in orientation since for them $\chi$ is the identity. This accords with expectations for chain complexes in general.

**Conclusions**

- Octonion co-ordinatization of the tangent bundle based on the four dimensional space-time manifold seems to chime in a surprising and pleasing way with a number of features of condensed state theory and in particular with the key construct of reciprocal space.
- It remains an open question as to whether the most appropriate scalars to use in octonion analysis are the real or the rational numbers.
- In their most general form the properties of three-space are expected to belong to projective geometry.
- A connection between the interior of the future light cone and a semi-algebraic set arising from octonion algebra is indicated.
- Analysis suggests that in classical electromagnetism the constitutive relations between the derived fields D and H and the fundamental fields E and B may not turn out to be unfathomably complicated after all.

# References

[1] Butcher P N and Cotter D 1990 *The Elements of Nonlinear Optics* (Cambridge: University Press) chapter 2

[2] Weyl H 2012 *Levels of Infinity* (New York: Dover) 203-216 Why is the World Four-Dimensional?

[3] Anderson P W 2013 Strange connections to strange metals *Physics Today* **66** April p9-10

[4] Schutz B F 1980 *Geometric Methods of Mathematical Physics* (Cambridge: University Press) p 37

[5] Gogberashvili M 2006 Octonionic Electrodynamics *J. Phys. A Math. Gen.* **39** 7099-7104

[6] Frankel T 2017 *The Geometry of Physics* (Cambridge: University Press: third edition) p200 remark #2

[7] Birss R R 1963 Macroscopic Symmetry in Space-Time *Rep. Prog. Phys.* **26** 307-360; table 2 following p319

[8] Born M and Wolf E 2017 *Principles of Optics* (Cambridge: University Press) p805

[9] Bunn C W 1946 *Chemical Crystallography* (Oxford: Clarendon Press) p85

[10] Potton R J 2014 Circular birefringence in crystal optics arXiv:1407.6797[physics.optics]

[11] Potton R J 2016 Discrete space-time symmetry, polarization eigenmodes and their degeneracies arXiv:1604.02320[physics.optics]

[12] Baez J C 2001 The Octonions *Bull. Of the Amer. Math Soc* **39** 145-205; arXiv:math /0105155 [math.RA] and references therein

[13] Okubo S 2005 *Introduction to Octonion and Other Non-Associative Algebras in Physics* (Cambridge: University Press) chapter 8

[14] Birss R R 1980 Multivector Analysis I: A Comparison with Tensor Algebra *Phys. Lett.* **78A** 223-226

[15] Dirac P  Projective Geometry, Origin of Quantum Equations *Audio recording made by John B. Hart*, *Boston University,* October 30 1972

[16] Adamo T 2015 General relativity as a two-dimensional CFT *Int J of Mod. Phys* **D 24** 1544024; arXiv:1505.05679v1[gr-qc]

[17] Coxeter H S M 1974 *Projective Geometry* (New York, Berlin Heidelberg: Springer-Verlag: second edition) p 104

[18] Weyl H 1973 *Classical Groups* (Princeton New Jersey: University Press: second edition) p165

[19] Dunnington G W 2004 *Carl Friedrich Gauss: Titan of Science* (Washington DC: Mathematical Association of America) p167

[20] The Open University 2006 SMT359 (Level 3 Electromagnetism) Book 2 *Electromagnetic fields* editor Freak S chapters 2 and 3